\documentclass{iau}
\usepackage{graphicx}
\usepackage{amsmath}
\usepackage{amssymb}
\usepackage{natbib}

\newcommand{\chan}{\textit{Chandra}}
\newcommand{\swift}{\textit{Swift}}
\newcommand{\rxte}{\textit{RXTE}}
\newcommand{\xmm}{\textit{XMM-Newton}}
\newcommand{\inte}{\textit{INTEGRAL}}

\newcommand{\rosat}{\textit{ROSAT}}

\newcommand{\suzaku}{\textit{Suzaku}}
\newcommand{\beppo}{\textit{BeppoSAX}}

\newcommand{\asca}{\textit{ASCA}}

\newcommand{\lum}{\mathrm{erg~s}^{-1}}
\newcommand{\flux}{\mathrm{erg~cm}^{-2}~\mathrm{s}^{-1}}

\newcommand{\nh}{\mathrm{cm}^{-2}}

\newcommand{\ks}{KS 1741--293}
\newcommand{\grs}{GRS 1741--2853}
\newcommand{\xte}{XTE J1701--462}
\newcommand{\xmmbron}{XMM~J174457--2850.3}
\newcommand{\grobron}{GRO J1744--28}

\def \mnras {\textit{MNRAS}}
\def \apj {\textit{ApJ}}
\def \apjs {\textit{ApJS}}

\def \aap {\textit{A\&A}}

\def \pasj {\textit{PASJ}}

\def \iaucirc {\textit{IAU Circ.}}

\title[IAUS290.~~\ks] 
{The transient neutron star X-ray binary \\ \ks\ in outburst and quiescence} 

\author[N. Degenaar \& R. Wijnands]  
{N. Degenaar$^{1}$\thanks{Hubble fellow}
 \and R. Wijnands$^2$}

\affiliation{$^1$University of Michigan, Dept. of Astronomy, 500
  Church St, Ann Arbor, MI 48109, USA \\
 email: {\tt degenaar@umich.edu} \\[\affilskip]
$^2$Astronomical Institute ``Anton Pannekoek", University of Amsterdam\\ Postbus 94249, 1090 GE Amsterdam, The Netherlands \\email: {\tt r.a.d.wijnands@uva.nl}}

\pubyear{2012}
\volume{290}  
\jname{\mbox{Feeding Compact Objects: Accretion on All Scales}}
\editors{C. M. Zhang, T. Belloni, M. Mendez \& S. N. Zhang (eds.)} 
\begin{document}

\maketitle

\begin{abstract}
\ks\ is a transient neutron star low-mass X-ray binary that is located at an angular distance of $\simeq$20$'$ from the Galactic center. We map out the historic activity of the source since its discovery in 1989, characterize its most recent X-ray outbursts observed with \swift\ (2007, 2008, 2010, and 2011), and discuss its quiescent X-ray properties using archival \chan\ data. \ks\ is frequently active, exhibiting outbursts that typically reach a 2--10 keV luminosity of $L_{\mathrm{X}} \simeq 10^{36}~(D/6.2~\mathrm{kpc})^2~\lum$ and last for several weeks--months. However, \swift\ also captured a very short and weak accretion outburst that had a duration of $\lesssim$4~days and did not reach above $L_{\mathrm{X}} \simeq 5 \times10^{34}~(D/6.2~\mathrm{kpc})^2~\lum$. The source is detected in quiescence with \chan\ at a 2--10 keV luminosity of $L_{\mathrm{X}} \simeq 2.5 \times10^{32}~(D/6.2~\mathrm{kpc})^2~\lum$.
\keywords{accretion, accretion disks, stars: neutron, X-rays: binaries, X-rays: individual (\ks)}
\end{abstract}


\firstsection 
\section{Introduction}

\ks\ is a transient neutron star low-mass X-ray binary (LMXB) that was discovered in 1989 August by the TTM onboard the KVANT module of the Mir space station \citep[][]{zand1991}. The detection of type-I X-ray bursts revealed its binary nature and testified to the presence of a neutron star. The source is located at an angular distance of $\simeq20'$ from the Galactic center, at an estimated distance of $D\simeq6.2$~kpc \citep[as inferred from type-I X-ray burst analysis;][]{chelovekov2011}. Throughout this work we assume a distance of $D\simeq6.2$~kpc when quoting X-ray luminosities.

\begin{figure}[tb]
\begin{center}
 \includegraphics[width=5.1in]{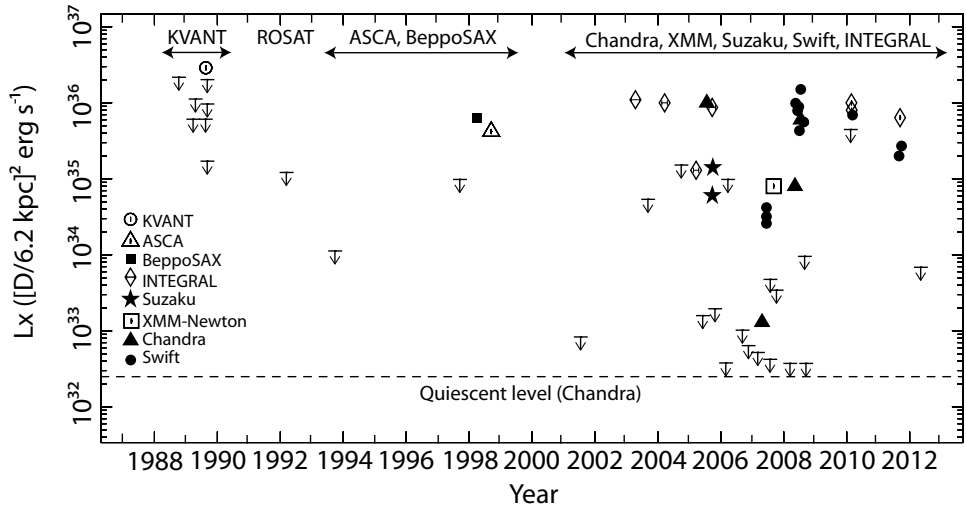} 
 \caption{{
 Long-term outburst history of \ks\ (2--10 keV; see text for references). 
 }}
 \label{fig:longlc}
\end{center}
\end{figure}

\section{The outburst history of \ks}\label{sec:hist}
We compiled a list of all (non-) detections of \ks\ reported in literature, and combined these with the analysis of archival data (Sections~\ref{sec:ob} and \ref{sec:quiescence}) to map out its historic  activity. Fluxes and upper limits were converted to the 2--10 keV energy band using \textsc{pimms}, assuming a power law spectrum with $N_{\mathrm{H}} = 21.5\times10^{22}~\nh$ and $\Gamma=2.1$. The long-term X-ray light curve as observed between 1989--2012 is shown in Figure~\ref{fig:longlc}.

After its discovery in 1989, the field around \ks\ was observed with \rosat\ in 1992, but the source was not detected \citep[Figure~\ref{fig:longlc};][]{sidoli01}. Renewed activity was seen in 1998 March and September by \beppo\ and \asca, which possibly covered the same outburst \citep[][]{zand1998,sidoli99,sakano02}. \inte\ detected it in outburst in 2003, 2004 and 2005 \citep[][]{cesare2007,kuulkers07,chelovekov2011}. Non-detections in between these epochs suggest that this were likely three separate outbursts (see Figure~\ref{fig:longlc}). The 2005 outburst was also observed by \chan\ and \suzaku\ \citep[][]{degenaar08_atel_gc_chan,yuasa2008}. 

Between 2006 and 2008, the source region was covered by monitoring campaigns of the Galactic center with \chan\ and \xmm\ \citep[][]{wijnands06,degenaar2012_gc}, and \swift\ \citep[][]{degenaar09_gc,degenaar2010_gc}. The source was found (weakly) active on three different epochs in 2007 \citep[Section~\ref{sec:ob};][]{degenaar2012_gc}, and a new outburst was observed in 2008 \citep[][]{degenaar2008,degenaar2012_gc}. \inte\ and \swift\ found the source active again in 2010 \citep[][]{chenevez2010} and in 2011 \citep[][]{barthelmy2011_ks1741,linares2011,chenevez2011_transients}. 

Only a few type-I X-ray bursts have been reported for \ks. KVANT detected two \citep[][]{zand1991}, six have been seen with \inte\ \citep[][]{cesare2007,chelovekov2011}, and recently one was picked up by \swift\ \citep[][]{barthelmy2011_ks1741,linares2011}. \rxte\ may have also detected one \citep[][]{galloway06}.

\section{Outbursts observed with \swift\ between 2007 and 2012}\label{sec:ob}
\begin{figure}[tb]
\begin{center}
 \includegraphics[width=5.0in]{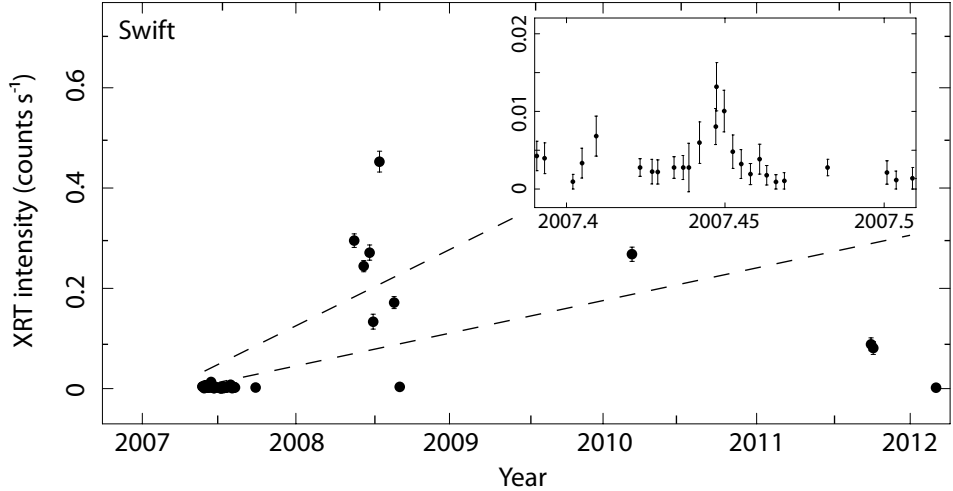} 
 \caption{{
 Long-term \swift/XRT light curve (binned per observation) showing three main outbursts of \ks\ (2008, 2010, and 2011) and one episode of low-level activity (2007). 
 }}
 \label{fig:swiftlc}
\end{center}
\end{figure}

We obtained all \swift/XRT data covering \ks\ from the public data archive. This encompasses 69 observations performed between 2007 May 5 and 2012 May 17. We extracted data products using the online XRT tools \citep[][]{evans09}. Figure~\ref{fig:swiftlc} displays the \swift/XRT light curve. It shows  three major outbursts in 2008, 2010, and 2011, as well as a short and weak episode of activity in 2007. We extracted spectra for these four different outbursts and fitted these simultaneously in \textsc{XSpec} to a simple absorbed power law model. The results of the spectral analysis are summarized in Table~\ref{tab:spec}.

The region around \ks\ was observed every few days by \swift\ between 2007 May 23 and August 9. The source is not detected during this epoch, except during three observations performed on June 13 and 14, when it displayed a 2--10 keV luminosity of $L_{\mathrm{X}}\simeq3\times10^{34}~\lum$. Non-detections in the preceding and subsequent observations (June 11 and 15) suggest that this episode of low-level activity had a duration of $\lesssim4$~days. 

\ks\ was detected at similarly low intensities at another two epochs in 2007. \xmm\ found the source active at $L_{\mathrm{X}}\simeq8\times10^{34}~\lum$ on 2007 September 6. Non-detections with \swift\ on August 9 and September 27 constrain the duration of this outburst to be $\lesssim49$~days. In an archival \chan\ observation performed on 2007 April 25, \ks\ is weakly detected at $L_{\mathrm{X}}\simeq1.3\times10^{33}~\lum$ (Table~\ref{tab:spec}). Non-detections with \chan\ on 2007 February 19 ($L_{\mathrm{X}}\lesssim5\times10^{32}~\lum$) and \swift\ between 2007 May 23 and June 10 ($L_{\mathrm{X}}\lesssim1\times10^{33}~\lum$) suggest that the activity lasted for $\lesssim$3~months. \ks\ was thus found active a factor $\simeq$10--100 above its quiescent level several times in 2007, but was not detected above $L_{\mathrm{X}}\simeq10^{35}~\lum$.

We obtained a series of \swift/XRT ToO observations of \ks\ between 2008 May 18 and September 4 to monitor the new outburst that was first detected by \chan\ on May 10 \citep[][]{degenaar08_atel_gc_chan,degenaar2008}. The source remained active for $\simeq$3~months at an average 2--10 intensity of $L_{\mathrm{X}}\simeq8\times10^{35}~\lum$ (Table~\ref{tab:spec}) till August 21. It was no longer detected on September 4 with an upper limit of $L_{\mathrm{X}}\lesssim9\times10^{33}~\lum$. In 2010 and 2011, \ks\ was again observed in outburst with \swift\ at an intensity of $L_{\mathrm{X}}\simeq(3-7)\times10^{35}~\lum$ (Table~\ref{tab:spec}, Figure~\ref{fig:swiftlc}).

\section{\ks\ in quiescence}\label{sec:quiescence}
To investigate the quiescent properties, we used archival \chan\ observations performed in 2001, 2006 and 2007 \citep[Obs IDs 2267, 7038, and 8459; from the campaign of][]{muno2009}. The source is not detected in the individual images, but clearly visible when the observations are combined ($\simeq$51~ks). We created a combined spectrum using the \textsc{ciao} tools, and fitted this simultaneously with the \swift\ outburst data (Table~\ref{tab:spec}). 

\ks\ is detected in quiescence at a 2--10 keV luminosity of $L_{\mathrm{X}}\simeq2.5\times10^{32}~\lum$, which is typical for quiescent neutron star LMXBs. 
The very large extinction in the direction of the source ($N_{\mathrm{H}} \simeq 2\times10^{23}~\nh$) completely obscures the thermal emission that is often detected for quiescent neutron star LMXBs (typically $kT_{\mathrm{bb}}$$\simeq$0.1--0.3~keV). 
Therefore, we only detect the hard (non-thermal) power law tail.

\begin{table}[tb]
\caption{Spectral properties of \ks\ in outburst and quiescence.}
\begin{center}
\begin{tabular}{c c c c c c}
\hline
Instr. & Date & State & $\Gamma$ & $F_{\mathrm{X}}$ ($\flux$) & $L_{\mathrm{X}}$ ($\lum$) \\
\hline
\chan & 2001/2006/2007 & quiescence &  $1.3^{+5.4}_{-1.3}$  & $(5.5 \pm 5.0) \times 10^{-14}$ & $(2.5 \pm 2.3) \times 10^{32}$ \\
\chan & 2007 Apr 25 & low activity  & $0.2^{+1.6}_{-0.2}$  & $(2.8 \pm 1.3) \times 10^{-13}$ & $(1.3 \pm 0.6) \times 10^{33}$  \\   
\swift & 2007 Jun 13--14 & low activity &  $0.8^{+2.4}_{-0.8}$  & $(6.2 \pm 1.4) \times 10^{-12}$ & $(2.9 \pm 0.6) \times 10^{34}$ \\
\swift & 2008 May 18--Aug 21 & outburst &  $ 2.1\pm 0.4$  & $(1.7 \pm 0.4) \times 10^{-10}$ & $(7.8 \pm 1.9) \times 10^{35}$  \\
\swift & 2010 Mar 10 & outburst & $2.2 \pm 0.6$  & $(1.5 \pm 0.5) \times 10^{-10}$ & $(6.9 \pm 2.3) \times 10^{35}$  \\
\swift & 2011 Sep 1--30 & outburst & $2.7 \pm 1.6$  & $(6.0 \pm 2.4) \times 10^{-11}$ & $(2.8 \pm 1.1) \times 10^{35}$ \\
\hline
\end{tabular}
\label{tab:spec}
Note. Quoted errors refer to $90\%$ confidence levels. A simultaneous fit to the spectral data resulted in $N_{\mathrm{H}}=(21.5\pm2.8) \times 10^{22}~\nh$ and $\chi_{\nu}^2 =1.1$ for 142 dof. $F_{\mathrm{X}}$ represents the unabsorbed 2--10 keV flux and $L_{\mathrm{X}}$ the corresponding luminosity assuming $D=6.2$~kpc. 
\end{center}
\end{table}

\section{Discussion}\label{sec:conclude}
\ks\ is a frequently active: between 1989 and 2012, the source exhibited at least 8 accretion outbursts that reached $L_{\mathrm{X}}\simeq10^{36}~\lum$ (2--10 keV) and had a duration of several weeks--months. This suggests a recurrence time of $\simeq$2~yr and a duty cycle of $\simeq$12.5\%. In addition to these main outbursts, we found indications of low-level accretion activity a factor of $\simeq$10--100 above the quiescent level of $L_{\mathrm{X}}\simeq2.5\times10^{32}~\lum$. Such peculiar behavior has now been observed for a number of transient neutron star LMXBs, such as \xmmbron, \grobron, \grs, and \xte\ \citep[][]{degenaar09_gc,degenaar2010_gc,degenaar2012_gc,fridriksson2011}.

~\\
\noindent {\bf Acknowledgements}\\
ND is supported by NASA through Hubble Postdoctoral Fellowship grant number HST-HF-51287.01-A from the Space Telescope Science Institute, and RW by a European Research Council starting grant. This work made use of \swift\ data supplied by the UK Swift Science Data Centre at the University of Leicester, and the \chan\ data archive.

\end{document}